\documentclass[english]{article}
\usepackage[T1]{fontenc}
\usepackage[latin1]{inputenc}
\usepackage{amssymb}
\usepackage{amsmath}
\usepackage{graphicx}
\makeatletter
\numberwithin{equation}{section}

\def\lsim{\mathrel{\rlap{\lower4pt\hbox{\hskip1pt$\sim$}}
    \raise1pt\hbox{$<$}}}                
\def\bes{\begin{equation}}
\def\beas{\begin{eqnarray}}
\def\ees{\end{equation}}
\def\eeas{\end{eqnarray}}
\def\be{\begin{equation}}
\def\bea{\begin{eqnarray}} 
\def\ee{\end{equation}}
\def\eea{\end{eqnarray}}

\def\bolde{\mathbf{e}}

\def\boldx{\mathbf{x}}
\def\Rhat{{\hat R}}
\def\deltatauhat{ \widehat{\Delta \tau} }
\def\deltathetahat{\widehat{\Delta \theta}}
\def\RHS{{\rm RHS}}
\def\tauhat{{\hat \tau}}
\def\tautilde{{\tilde \tau}}

\def\khat{\mathbf{ \hat k}}

\newcommand{\appendices}{
  \renewcommand{\thesection}{\Alph{section}}
  \setcounter{section}{0}
  \newpage
}

\usepackage{babel}
\makeatother
\begin{document}

\begin{titlepage}
\nopagebreak

\title{
Local bulk S-matrix elements 
and CFT singularities} 

\vfill
\author{Michael Gary,$^{a,}$\footnote{mgary@physics.ucsb.edu}\ \ Steven B. Giddings,$^{a,}$\footnote{giddings@physics.ucsb.edu}\ \  and Joao Penedones$^{b,}$\footnote{penedon@kitp.ucsb.edu}
}


\maketitle

\vskip 0.5cm

{\it  $^{a}$ Department of Physics, 
University of California, Santa Barbara, CA 93106, and PH-TH, CERN, Geneva, Switzerland}

{\it $^{b}$ Kavli Institute for Theoretical Physics,
University of California, Santa Barbara, CA 93106-4030, USA}
\vfill
\begin{abstract}
We give a procedure for deriving certain bulk S-matrix elements from corresponding boundary correlators.  These are computed in the plane wave limit, via an explicit construction of certain boundary sources that give 
bulk wavepackets.  A critical role is played by  a specific singular behavior of the lorentzian boundary correlators.  It is shown in examples how correlators derived from the bulk supergravity exhibit  the appropriate singular structure, and reproduce the corresponding S-matrix elements.  This construction thus provides a nontrivial test for whether a given boundary conformal field theory can reproduce bulk physics, and where it does, supplies a prescription to extract bulk S-matrix elements in the plane wave limit.

 \end{abstract}
 \vskip.4in
 
\noindent
CERN-PH-TH/2009-035\hfill \\  
NSF-KITP-09-35\hfill \\  
\vfill
\end{titlepage}

\section{Introduction}

Since the AdS/CFT correspondence was proposed\cite{juan}, an important question has been how it can be read to determine properties of the bulk theory from CFT quantities.  There has been an enormous amount of investigation of various aspects of the boundary behavior that are implied by different features of the bulk theory, but there has been comparatively little investigation of the problem of extracting bulk behavior from the boundary theory.  This problem, which might be referred to as ``decoding the hologram,'' has remained challenging. Of course, one of the purported miracles of the correspondence is precisely related to this question, that of how a higher-dimensional theory could be fully encoded in the lower-dimensional one.  

One of the challenges is to find sufficiently sharply refined boundary quantities that would be sensitive to fine-grained bulk detail.  We will particularly focus on the question of whether and how the bulk S-matrix could be extracted from the boundary theory.  Proposals for a prescription to do so from boundary correlators were outlined in \cite{Polchinski:1999ry,Susskind:1998vk}.  However, there are considerable subtleties in implementing such proposals, described for example in \cite{FSS}.  In particular, generic boundary data that might be used to specify incoming states, corresponding to non-normalizable behavior of states near the boundary, produces divergences that obscure physics at scales short compared to the AdS scale, $R$.  

This suggests that one consider more specialized boundary sources.  In order to localize at scales smaller than $R$ in the bulk, one expects to need a construction of bulk wavepackets that do so. However, in order to avoid the divergences from the non-normalizable behavior, one expects  that such  data should have compact support on the boundary.

The present paper will propose a class of sources and corresponding bulk wavepackets, which appear to strike an optimal balance between these criteria.  Specifically, they have compact boundary support and are also taken to have high-frequency modulation in order to maximize localization properties.  We then use these wavepackets to see how  bulk S-matrix elements in the plane wave limit can be extracted from a candidate class of boundary correlators, in a certain scaling limit.\footnote{This scaling limit is closely related to the proposal in \cite{Polchinski:1999ry}. }  This can happen only if the correlators have a certain singularity structure, which, we will explain, is necessary for producing the correct bulk kinematics.  
This singularity is only visible in the lorentzian regime and can be reached by a specific analytic continuation of the euclidean correlator, similar to the one considered in  \cite{Cornalba:2006xk,Cornalba:2006xm,Cornalba:2007zb,Cornalba:2007fs,Cornalba:2008qf} in the context of the Regge limit of CFT correlators and eikonal scattering in AdS.
While the correlators we consider are those arising via the GKPW prescription \cite{GKP,Witten} from the bulk supergravity, and thus do not arise directly from a boundary CFT, this construction provides an important test in principle, elucidating  necessary CFT structure to encode certain local bulk dynamics.  Thus, we suggest it supplies a piece of the answer to the question of how one might ``decode the hologram.''

In outline, the next section will give our explicit boundary sources and will investigate some bulk properties of the corresponding wavepackets.  Next, section three, first from the bulk perspective, explains how such wavepackets could be tuned via an appropriate scaling limit to extract S-matrix elements for plane waves,  from a simultaneous $R\rightarrow\infty$ and large-wavepacket limit.  Then, a corresponding discussion is given on the boundary side, where one finds that the bulk kinematics of the momentum-conserving delta function can be encoded in a certain boundary singularity structure and that, for CFTs with this structure, the reduced transition matrix element could be read off from the coefficient of the singularity.  Finally, section four shows that certain proposed CFT correlators, derived in other works from bulk supergravity, reproduce the correct singularity structure and, via our prescription,  the correct reduced transition matrix elements.  Two appendices contain technical details.

\section{Wavepackets in AdS}
\label{wavepacket}

Our starting point will be to specify boundary data that constructs wavepackets, whose scattering we will then study.  Our goal is to have these wavepackets sufficiently localized that they scatter only within a region small as compared to the AdS radius.  

\subsection{Geometry and coordinates}
\begin{figure}
\centering
\includegraphics[width=11cm]{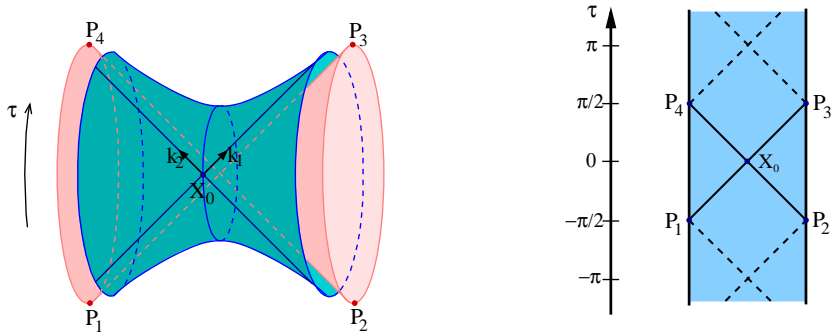}
\caption{AdS$_2$ is shown in blue. The revolution axis of the hyperboloid corresponds to the spacelike direction of $\mathbb{R}^{2,1}$ and the transverse plane is timelike. Global time is the angular coordinate in this plane.
The point $X_0$ is a reference point in AdS$_2$. 
The null momenta $k_1$ and $k_2$ live in the tangent space to AdS at  $X_0$. 
The boundary sources are supported in the neighborhood of the boundary points $P_i$. 
On the right, we show the universal cover of AdS$_2$ conformally compactified.  }\label{hyperboloid}
\end{figure}

We begin by reviewing some basics of AdS geometry.\footnote{See \cite{FSS,Penedones:2007ns} for more description of the geometry and the relation to the embedding space.} 
AdS$_{d+1}$ may be thought of as a hyperboloid $(X^M)^2=-R^2$ in $\mathbb{R}^{2,d}$. AdS$_2$ embedded in $\mathbb{R}^{2,1}$ is shown in figure \ref{hyperboloid}. The point $X_0=(R,0,0,...)$, which we will take as the point about which our wavepackets will intersect, is also shown. 

For present purposes we will work on the cover of AdS.  We will parameterize this in terms of global coordinates $(\tau,\rho,\bolde)$, with $\bolde$  a $d$-dimensional unit vector on $S^{d-1}$; these are related to embedding coordinates by
\begin{equation}
X=\frac{R}{\cos\rho}(\cos\tau,\sin\tau,\sin\rho\,\mathbf{e}). \label{globalcoords}
\end{equation}
Then the metric takes the form
\bes
\label{globalmet}
ds^2={R^2\over \cos^2\rho}\left(-d\tau^2 + d\rho^2 + \sin^2\rho\, d\bolde^2\right)\ .
\ees
Points on the boundary of AdS, $\rho\rightarrow \pi/2$, naturally map to the corresponding null rays in the embedding space, 
\begin{equation}
\label{boundcoords}
P=(\cos\tau,\sin\tau,\bolde)\ .
\end{equation}

The flat space limit of the vicinity of the point  $x_0$, which corresponds to $X_0$, is taken by defining coordinates $t=R\tau$, $r=R\rho$.  Then, the metric (\ref{globalmet}) becomes
\begin{equation} ds^2=\frac{1}{\cos^2\frac{r}{R}}\left[ -d t^2+ dr^2 +R^2 \sin^2 \frac{r}{R}  d\mathbf{e}^2 \right]\ .\end{equation} 
For $t,r\ll R$, this approximates the flat metric.  The corresponding neighborhood in the embedding space is given by 
\begin{equation}
X\approx (R,t,\mathbf{x}) \ ,\label{centralcoords}
\end{equation}
where $\mathbf{x}=r\mathbf{e} \in \mathbb{R}^d$.  Thus AdS$_{d+1}$ is well-approximated by its tangent space $\mathbb{M}^{d+1}$,
which is simply the subspace of $ \mathbb{R}^{2,d} $ orthogonal to $ X_0=(R,0,\dots,0)$.

%

\subsection{Wavepackets}

We wish to describe boundary sources $\Phi$ that construct bulk wavepackets $\Psi$ that have appropriate properties, such as localization, {\it etc.}  These will be related through the bulk-boundary propagator, 
\begin{equation}
G_{B\partial}(b,x)= {C_\Delta \over R^{(d-1)/2}} {1\over (-2P\cdot X/R+i\epsilon)^{\Delta}}\ ,
\end{equation}
where $b=(\tau,\bolde)$ is a boundary point, the product in the denominator is formed between the embedding-space quantities corresponding to $b$ and $x$, via (\ref{globalcoords}) and (\ref{boundcoords}), $\Delta$ is the conformal dimension, and the $i\epsilon$ prescription is that appropriate to the neighborhood of the point $x_0$.\footnote{For more discussion of the  $i\epsilon$ prescription on the universal cover of AdS see \cite{Dullemond:1984bc,Penedones:2007ns}.}  
 In our conventions, which include powers of $R$ to yield appropriate bulk dimensions for bosonic fields, the constant $C_{\Delta}$ is given by
\bes
C_{\Delta} =\frac{\Gamma(\Delta)}{2 \pi^{\frac{d}{2}}\Gamma\left(\Delta-\frac{d}{2}+1\right)}\ .
\ees
The bulk wavepacket will be given by
\begin{equation}
\label{bulkboundary}
\Psi(x)=\int_{\partial AdS} d\tau d\bolde \Phi(b)G_{B\partial}(b,x)\ .
\end{equation}

It is important that the boundary source $\Phi$ be a smooth, compactly supported function. The requirement of compact support will become necessary when considering multiple boundary sources, as there are divergences when sources overlap, as emphasized in \cite{FSS}. From the AdS side, this can be understood by noting that if there are multiple overlapping non-normalizable modes, integrals over bulk interaction points will not converge. It is also possible to understand the divergence from the CFT perspective, where it arises when two CFT operators approach the same point. The smoothness requirement is useful in ensuring the approximations made later in the paper are well controlled, forcing the Fourier transform $\hat{\Phi}$ to fall faster than any power at high frequencies.\footnote{A smooth, compactly supported function is an example of a Schwartz function. The Fourier transform acts as an endomorphism on the space of Schwartz functions \cite{Reed:1980}.}

We shall localize our source around the boundary point $b_0=(\tau_0,\bolde_0)$.  We also introduce explicit frequency dependence, in order to produce a bulk wavepacket with frequencies near $\omega$.  Thus, we consider a boundary source of the form
\begin{equation}
\Phi_{(\omega,\tau_0,\mathbf{e}_0)} (b)=  
e^{-i\omega R(\tau-\tau_0)} 
L\left(\frac{\tau-\tau_0}{\Delta\tau}\right) L\left(\frac{\theta}{\Delta\theta} \right) \label{source}
\end{equation}
where  $L$ is a $C^\infty$ function with $L(0)=1$ and with compact support of width $\sim 1$ about the origin, 
\bes
\cos\theta = \bolde\cdot\bolde_0\ ,
\ees
and $\Delta \tau$ and $\Delta \theta$ thus give the widths of the wavepacket on the boundary.
In order to achieve the desired localization, we take 
\begin{equation}
\frac{1}{\omega R} \ll \Delta\tau, \Delta\theta \ll 1\ . \label{boundcond}
\end{equation}

\subsection{Wave packet in the interaction region}
\label{wavepackint}

We next examine the form of the bulk wavepacket corresponding to the source (\ref{source}) in the vicinity of the point $x_0$, which we would like to be the interaction region between such wavepackets.  
In order that the wavepacket pass through this region, we take $\tau_0=-\pi/2$.  Near $x_0$, and for small  ${\tilde \tau} =\tau+\pi/2$, (\ref{boundcoords}) and (\ref{centralcoords}) give
\bes
-P\cdot X\approx  R\cos\tau + t\sin\tau - \bolde\cdot \boldx \approx R\tautilde -t - \bolde\cdot \boldx\ .
\ees
Substituting this into equation (\ref{bulkboundary}), we find
\bes
\Psi_{(\omega,\mathbf{e}_0)}(x)\approx C_{\Delta}
 R^{\Delta -(d-1)/2}
\int d\tautilde \int d\mathbf{e}
{e^{-i\omega R\tautilde} L\left(\frac{\tautilde}{\Delta\tau}\right) L\left(\frac{\theta}{\Delta\theta} \right)\over \left(2(R\tautilde -t - \bolde\cdot \boldx) +i\epsilon\right)^\Delta             }\ ,
\ees
and the substitution and $\chi = \omega(R\tautilde -t-\bolde\cdot \boldx)$ gives
\begin{eqnarray}
\Psi_{(\omega,\mathbf{e}_0)}(x) &\approx& C_{\Delta}  \omega^{\Delta-1}R^{\Delta -(d+1)/2}
e^{-i\omega t}\notag
\\ 
&&\int d\chi \int d\bolde {e^{-i\chi-i \omega \bolde\cdot\boldx } L\left(\frac{\chi+\omega(t + \bolde\cdot \boldx)}{\omega R\Delta\tau}\right)
L\left(\frac{\theta}{\Delta\theta} \right)\over (2\chi +i\epsilon)^\Delta} \ .\notag\\
\end{eqnarray}

For large $\omega R \Delta \tau$, the dominant contribution to the integral over $\chi$ comes from the singularity at 
$\ \chi= 0$;  subleading contributions are given by the series expansion in $\chi$ of the source near the singularity, 
and hence are supressed by powers of $\frac{1}{\omega R\Delta\tau}$. Thus, 
\begin{equation}
\label{bulkpack}
\Psi_{(\omega,\mathbf{e}_0)}(x)\approx \mathcal{D}_\Delta e^{-i\omega t}  
 \omega^{\Delta-1}R^{\Delta -(d+1)/2}
\int d\mathbf{e} e^{-i\omega\mathbf{e}\cdot\mathbf{x}}
L\left(\frac{t +\mathbf{e}\cdot \mathbf{x} }{R\Delta\tau}\right)L\left(  {\theta \over \Delta \theta} \right)
\end{equation}
where
\begin{equation}
\label{DDelta}
\mathcal{D}_\Delta=\frac{2\pi C_{\Delta} e^{-i\pi\Delta/2}}{2^\Delta\Gamma(\Delta)}.
\end{equation}

For $t,r\ll R\Delta \tau$, the first $L\approx1$, and the integral over angles gives the Fourier transform of the angular source.    Thus, in the limit of small $\Delta \theta$, we find 
\bes
\Psi_{(\omega,\mathbf{e}_0)}(x)\approx 
\Psi_{(\omega,\mathbf{e}_0)}(0)e^{ik\cdot x } \label{planewave}
\ees
where $k=\omega(1,-\bolde_0)$.  The coefficient $\Psi_{(\omega,\mathbf{e}_0)}(0)$ is easily evaluated, to give
\bes
\Psi_{(\omega,\mathbf{e}_0)}(0)= \mathcal{D}_\Delta  \omega^{\Delta-1}R^{\Delta -(d+1)/2} \int d\mathbf{e} L\left(  {\theta \over \Delta \theta} \right)\ .
\ees
In the limit of small $\Delta \theta$, 
\bes
\int d\mathbf{e} L\left(  {\theta \over \Delta \theta} \right) = (\Delta\theta)^{d-1} {\tilde L}_{d-1}\ ,
\ees
with
\bes
{\tilde L}_{d-1} = \int d^{d-1}\kappa L(\kappa) = \Omega_{d-2}\int_0^\infty \kappa^{d-2} d\kappa L(\kappa) \ 
\ees
and $\Omega_{d-2}$ the volume of unit $S^{d-2}$.

Outside of the small $\Delta\theta$ limit, we see that the function $\Psi_{(\omega,\mathbf{e}_0)}(x)$ is a wavepacket with characteristic widths given by
\bes
\label{bulksize}
\Delta t = R\Delta \tau\quad,\quad \Delta x_\perp = {1\over \omega \Delta \theta}
\ees 
in the longitudinal  direction $t+\bolde_0\cdot \boldx$ and in the transverse directions, respectively.
If we wish to have a wave packet that looks approximately like a plane wave near $x_0$, but is well localized at short distances as compared to $R$, we require
\begin{equation}
\label{bulkcond}
\frac{1}{\omega}\ll \Delta t, \Delta x_\perp \ll R\ ,
\end{equation}
which, using (\ref{bulksize}), is equivalent to (\ref{boundcond}).

\section{Flat space S-matrix elements from CFT correlators}
\label{flatS}

Our goal will be to establish a relation between the CFT correlators and elements of the S-matrix of the dual string theory in the flat (Minkowski) limit.  We will do so via a limiting procedure similar to that proposed in \cite{Polchinski:1999ry,Susskind:1998vk}. Specifically, we will scatter four of the wavepackets we have described, in the vicinity of the point $x_0$, adjusting them such that their typical widths remain less than the AdS radius.  This is then expected to confine the interactions to the (almost) flat neighborhood of   
this point and, under the limit of infinite AdS radius, allows one to extract flat space S-matrix elements.
In particular, we shall focus on $2\to 2$ elastic scattering with $\Delta_3=\Delta_1$ and  $\Delta_4=\Delta_2$.  

Since this construction is based on choosing specific boundary sources integrated against a CFT correlation function, this then exhibits the appropriate limit of such correlators to be taken if bulk S-matrix elements are to be extracted.  We emphasize that we will ultimately work with correlation functions that are derived from the bulk supergravity (or string) Feynman rules, as in \cite{GKP,Witten}.  By showing how to isolate the needed behavior from such correlators that have a local bulk origin, we thus provide a test that can be applied to a true boundary conformal field theory, such as ${\cal N}=4$ super-Yang Mills, to see whether it has the appropriate structure to correspond to a local bulk theory.

\subsection{Bulk construction}

We work in the vicinity of $x_0$, approximately parameterized by (\ref{centralcoords}).  We will use wavepackets as given by (\ref{source}), with $(d+1)$-dimensional momenta
\bes
\label{STk}
k_i = (\omega_i, \mathbf{k}_i)\ .
\ees
 Our convention is that all momenta flow {\it into} the corresponding diagram, and in particular, $\omega_{3,4}$ are negative.  It will also be useful to define the corresponding (inward pointing) unit vectors,
 \bes
 \label{Sk}
 \mathbf{k}_i = |\omega_i|  { \mathbf{ \hat k}_i}\ .
 \ees

We thus take wavepackets (\ref{source}) with $\tau_{1,2}=-\pi/2$, and $\tau_{3,4}=\pi/2$, defined in terms of the angles 
\bes
\label{thetadef}
\cos\theta_i = -\bolde_i \cdot { \mathbf{ \hat k}_i}\ .
\ees
These produce bulk wavefunctions $\Psi_{k_i}(x)$ with behavior as described in section \ref{wavepackint}.
The scattering amplitude between these wave functions reads
\begin{equation} 
\label{Corrfcn}
\int_{{\rm AdS}} \prod_{i=1}^4 d x_i \Psi_{k_i}(x_i) G(x_1,\dots,x_4)
\end{equation} 
where $G(x_1,\dots,x_4)$ is the amputated bulk Green's function.

We shall take the flat space limit and the plane wave limit together, as in \cite{Polchinski:1999ry}. Specifically, we  introduce a dimensionless scaling parameter $\eta$ and define
\begin{equation} 
\label{scalelim}
R= \eta^2 \Rhat \ ,\ \ \ \ \ \ \ \ \Delta \tau = \eta^{-1}\deltatauhat\ , \ \ \ \ \ \ \ \ \Delta \theta  =\eta^{-1}\deltathetahat\ .
\end{equation} 
We then take the limit of large $\eta$, holding the $\omega_i$ and hatted quantities fixed.
 The conditions (\ref{boundcond}) and  (\ref{bulkcond}) are automatically satisfied due to the strong ordering $1 \ll \eta \ll \eta^2 $.
In this limit, the curvature corrections become small because the range of the wave packets scales with $\eta$ and the AdS radius of curvature scales with $\eta^2$. In the flat region, the wavepackets take the form
\bes
\Psi_{k_i}(x)\approx e^{ik_i\cdot x} F_i(x)\ ,
\ees
where the envelope $F(x)$ is given by (\ref{bulkpack}), and becomes nearly constant.

In the absence of IR divergences or other subtleties, we thus expect that in (\ref{Corrfcn}) we can replace the AdS Green function $G$ by the corresponding flat-space Green function, which we write in the form
\begin{equation} 
 G(x_1,\dots,x_4)=i \int_{\mathbb{M}^{d+1}} \prod_{i=1}^4 \frac{dk'_i}{(2\pi)^{d+1}} e^{-i  k'_i\cdot x_i}
 \mathcal{M}(k'_1, \dots, k'_4 )\ .
 \end{equation} 
The scattering amplitude (\ref{Corrfcn}) then becomes
\begin{equation} 
 \int_{\mathbb{M}^{d+1}}\prod_{i=1}^4  \frac{d k'_i}{(2\pi)^{d+1}} \hat{F}_i(k_i-k_i') 
\mathcal{M}(k_1',k_2',k_3',k_4')
\label{FMeqn}
\end{equation} 
where $\hat{F}_i$ is the Fourier transform of $F_i$. In the limit $\eta \to \infty$, the support of $\hat{F}_i$ gets localized at $k_i-k_i'\sim 1/\eta \to 0$.  In particular, one finds that 
\bes
\hat{F}_i(k_i-k_i')\rightarrow (2\pi)^{d+1}\delta^{d+1}(k_i-k'_i) \Psi_{k_i}(0)\ .
\label{Flim}
\ees

Of course, $ \mathcal{M}$ is directly related to the flat S-matrix, modulo the usual subtleties of LSZ, etc.  In particular,  the S-matrix has the form ${\cal S} = 1+i {\cal T}$, and for two-particle scattering between plane-wave states,
\bes
\langle k_3,k_4| {\cal T}|k_1,k_2\rangle = (2\pi)^{d+1} \delta^{d+1}\left(\sum_i k_i\right) T(s,t)\ ,
\ees
where one typically defines the Mandelstam invariants
\bes
\label{mandelstam}
s= -(k_1+k_2)^2\quad,\quad t= -(k_1+k_3)^2\quad,\quad u=-(k_1+k_4)^2\ ,
\ees
and $s+t+u=0$ for massless particles.
The scattering angle $\Theta$ is given by 
\bes
\sin^2 \frac{\Theta}{2} =-\frac{t}{s} \ ,\ \ \ \ \ \ \  \ \ \  \cos^2 \frac{\Theta}{2} =-\frac{u}{s}\ ,
\ees
and $s$ is the square of the center-of-mass energy.
Specifically, there is a direct contribution corresponding to the ``one'' in $\cal S$, which would arise from disconnected diagrams, and connected diagrams produce $T$.  Focussing on the latter, we expect ${\cal T} = {\cal M}$, and thus combining (\ref{FMeqn}), (\ref{Flim}), that 
\begin{equation}
 i (2\pi)^{d+1} \delta^{d+1} \left( \sum k_i \right)T(s,t)
=
\lim_{\eta \to \infty}  
   \int\prod_{i=1}^4 db_i {\Phi_{k_i} (b_i) \over\Psi_{k_i}(0)}   A_{CFT}(b_1,\cdots,b_4)\,. 
\label{mainformula}
\ees
Thus, we expect to be able to derive such elements of the S-matrix, corresponding to plane wave external states, from this limiting procedure.

\subsection{CFT construction}
\label{CFTconstruction}
The goal of this subsection is to see this procedure work, directly at the level of the CFT.  The reason for this is two-fold.  First, it may strike one that we could have been incautious in the limiting procedures of the preceding subsection.  Secondly, the formula 
(\ref{mainformula}) has on its left hand side basic features characteristic of a bulk local theory.  We would like to understand how these are reproduced by the CFT quantities, on the right hand side.  In fact, this could be viewed as providing a non-trivial test for CFTs, to diagnose whether they could correspond to a local (or approximately local) bulk theory, and moreover provide an important part of the key to decoding bulk local behavior from CFT correlators.

Specifically, combining (\ref{source}) and (\ref{mainformula}), together with the scaling limit (\ref{scalelim}), 
our conjecture is that the S-matrix elements are given by
\begin{eqnarray}
i (2\pi)^{d+1} \delta^{d+1} \left( \sum k_i \right)T(s,t)
&=&\lim_{\eta\rightarrow\infty}
 \int \prod_{i=1}^4\Bigl[ db_i N_i   e^{-i\omega_i \Rhat(\tau_i-\tau_{i0} )\eta^2}\notag\\
 &&   L\left({\eta (\tau_i-\tau_{i0} )\over \deltatauhat} \right) L\left({\eta\theta_i\over \deltathetahat}\right)
      \Bigr] A_{CFT}(b_i)\ ,
      \label{mainformulaa}
\end{eqnarray}
where
\bes
N_i = {1\over \eta^{2(\Delta_i-d)}} {1\over \mathcal{D}_{\Delta_i} \tilde{L}_{d-1}|\omega_i|^{\Delta_i -1} \Rhat^{\Delta_i -(d +1)/2} \deltathetahat^{d-1}}\ .
\ees
We wish to see how, for a given CFT, the expression on the right hand side produces the left hand side.

The form of the CFT four-point function is highly constrained by conformal invariance.  Let $P_i = P(b_i)$ be given by (\ref{boundcoords}); then
\begin{equation} 
\label{ACFTdef}
A_{CFT}(b_i)=
\frac{C_{\Delta_1} C_{\Delta_2} }
{(-2 P_1 \cdot P_3+i\epsilon)^{\Delta_1} (- 2 P_2\cdot P_4 +i\epsilon)^{\Delta_2} }
 \mathcal{A} (z,\bar{z})\, ,
\end{equation} 
where we chose the normalization so that $\mathcal{A}=1$ corresponds to the disconnected contribution  and
$z$ and $\bar{z}$ are defined in terms of the cross ratios
\begin{equation} 
z\bar{z}=\frac{(P_{1} \cdot P_{3})(P_{2} \cdot P_{4})}
{(P_{1} \cdot P_{2})(P_{3} \cdot P_{4})}\,,
\end{equation} 
and
\begin{equation} 
(1-z)(1-\bar{z})=\frac{(P_{1} \cdot P_{4})(P_{2} \cdot P_{3})}
{(P_{1} \cdot P_{2})(P_{3} \cdot P_{4})}\,.
\end{equation} 
A first check of (\ref{mainformula}) is that the RHS is zero for generic $k_i$'s not summing to zero.
In this case, we are probing the four point function at  generic values of the cross ratios where the function is regular.
Therefore, the $\tau_i$ integrals over the boundary will generate a Fourier transform of a smooth compact support function. 
In the limit $\eta \to \infty$, the frequency scales with $\eta^2$ and the width scales with $1/\eta$.  
Thus the final result is zero as expected.

We expect the needed delta function to arise from singular behavior of $\mathcal{A} (z,\bar{z})$.  The wavepackets in (\ref{mainformulaa}) force the $b_i$ to be in the vicinity of $(\tau_{i0}, - \mathbf{\hat{k}}_i)$, as seen from equations (\ref{STk}-\ref{thetadef}).
To exhibit the singular behavior, we let $\tau_{i0}=-\pi/2$ for $i=1,2$, and $\tau_{i0}=\pi/2$ for $i=3,4$.   Let us introduce new parameters $\rho$, $\sigma$ through  $z=\sigma e^{-\rho}$ and  $\bar{z}=\sigma e^{\rho}$; these can be shown to be given by
\begin{equation} 
\sigma^2=\frac{(P_{1} \cdot P_{3})(P_{2} \cdot P_{4})}
{(P_{1} \cdot P_{2})(P_{3} \cdot P_{4})}\,,
\end{equation} 
and
\bes
\label{rhodef}
\sinh^2 \rho= \frac{{\rm Det}(P_{i} \cdot P_{j})}
{  4 (P_{1} \cdot P_{3})(P_{2} \cdot P_{4})(P_{1} \cdot P_{2})(P_{3} \cdot P_{4})}\ .
\ees
We recognize in this last expression the Graham determinant of the null vectors $P_i$.  
For momentum-conserving $k$'s, we then find that $P(\tau_{i0}, - \mathbf{\hat{k}}_i)$ yield $\rho=0$ -- this follows from linear dependence of the vectors $P_i$.
We conclude that, for momentum-conserving $k$'s, the $\eta \to \infty$ limit in (\ref{mainformula}) is probing the $\bar{z}\approx z$ region of the correlator.

The reduced amplitude  $\mathcal{A} (z,\bar{z})$ is in general a dimensionless function of the cross ratios.
We have found that to produce the correct bulk structure, 
$\mathcal{A}$ must diverge  in the kinematical limit $ \bar{z} \to z$.
We shall focus on describing tree level interactions in $(d+1)$-dimensional spacetime controlled by the coupling constant $g$.
Then, the amplitude $\mathcal{A}$ will be proportional to the dimensionless factor $g^2 R^{5-d-2j}$,
where $j$ is fixed by dimensional analysis. This interaction corresponds to a flat space matrix element of the form
$i g^2 s^{j-1} \mathcal{M}(\Theta)$, where $\Theta$ is the scattering angle. In the particular case where the external scalars are minimally coupled to an exchanged particle, 
$j$ is the spin of the exchanged particle.  Thus, we expect an amplitude that behaves as 
\begin{equation}
\mathcal{A} (z,\bar{z})\approx g^2 R^{5-d-2j} \frac{\mathcal{F}(\sigma)}{(-\rho^2)^{\beta}} \ . \label{limA}
\end{equation}
in the vicinity $z\approx \bar{z}$.
For now, we assume this generic power law divergence but, below, we shall be able to fix the exponent $\beta$ by requiring appropriate scaling.
Moreover,  in section \ref{examples} we will examine some specific amplitudes (computed via the bulk supergravity)
and confirm that they exhibit such singularities.
We will also examine the kinematical origin of the singularity at $\rho=0$ later in this section.

The full delta function on momenta follows from the detailed structure of this singularity.  We begin by defining new variables ${\hat \tau}_i= (\tau_i -\tau_{i0}) \eta^2$, in terms of which the RHS of  (\ref{mainformulaa}) becomes
\bes
\RHS= \lim_{\eta\rightarrow\infty}{1\over \eta^8} \int \prod_{i=1}^4 d{\hat \tau}_i d\bolde_i N_i  e^{-i\omega_i\Rhat {\hat\tau}_i} 
L\left({{\hat \tau}_i\over\eta \deltatauhat} \right)L\left({\eta\theta_i\over \deltathetahat}\right) A_{CFT}(\tau_{i0}+{ {\hat \tau}_i\over \eta^2},\bolde_i)
\ees
In the limit $\eta\rightarrow\infty$, the integral becomes very peaked in $\theta_i$, {\rm i.e.} at the points $\bolde_i = -\mathbf{ \hat k}_i$.  Moreover, the distribution $L({{\hat \tau}_i/\eta \deltatauhat})$ becomes very flat, as compared to the variation in the exponential.  We thus replace it by its value at zero, $L(0)=1$.  The result is that 
\bes
\label{intermexp}
\RHS= \lim_{\eta\rightarrow\infty}{{\cal L}\over \eta^8}  \int\prod_i  d{\hat \tau}_i e^{-i\sum_i \omega_i \Rhat {\hat\tau}_i} A_{CFT}(\tau_{i0}+{ {\hat \tau}_i\over \eta^2}, -\mathbf{ \hat k}_i)\ 
\ees
where 
\bes
{\cal L}=  \prod_i N_i \left(\frac{\deltathetahat}{\eta}\right)^{d-1}{\tilde L}_{d-1}=
\prod_i
 {1\over \eta^{2\Delta_i-d-1}} {1\over \mathcal{D}_{\Delta_i} |\omega_i|^{\Delta_i -1} \Rhat^{\Delta_i -(d +1)/2} }
\\ .
\ees

A non-vanishing result comes from the singularity at ${\hat \tau}_i=0$, which for momentum-conserving $k_i$ produces the singularity at $\rho=0$.  The delta function follows by examining perturbations of this singularity as the momenta are varied away from conserved values. We first examine the contributions of the Graham determinant in (\ref{rhodef}).  With $b_i=(\tau_{0i}+{\hat \tau}_i/\eta^2,-\mathbf{ \hat k}_i)$, we find via (\ref{boundcoords})
\bes
P_i\cdot P_j = \pm\left( {1\over 2} {{\hat\tau}_{ij}^2\over \eta^4 } + {k_i\cdot k_j\over \omega_i\omega_j}\right) +{\cal O}[({\hat\tau}/\eta^2)^4]
\ees 
with ${\hat\tau}_{ij} = {\hat\tau}_i -{\hat\tau}_j$, and with plus sign for $(i,j)=(1,2)$ or $(3,4)$, and minus otherwise.
Thus, the determinant yields
\bes
\det(P_i\cdot P_j) = \det\left({k_i\cdot k_j\over \omega_i\omega_j}\right) + {\cal O}(\tauhat^2)\ .
\ees

While it should be possible to derive expressions in a general frame, we find it convenient to pick a particular frame to evaluate the quantities entering the correlator.  We do this using the isometry group of AdS, $SO(d,2)$.  This contracts to the flat Poincare group, so that such transformations can be used to pick particular Lorentz frames.  

In making coordinate choices, we also note that $A_{CFT}$ is invariant under translations of $\tau$.  This means that we can take $\tauhat_i\rightarrow \tauhat_i -\tauhat_1$, and eliminate $\tauhat_1$ from the correlator.  The integral over ${\hat \tau}_1$ then gives $2\pi\delta(\Rhat \sum_i\omega_i)$.  Then, let us choose $\khat_2 = -\khat_1$, as part of going to the center of mass frame.  Moreover, in general $\khat_1$ and $\khat_3$ define a plane; let $\khat_{4,\perp}$ be the projection of $\khat_4$ perpendicular to that plane.  Also, define $\cos\vartheta_3= -\khat_3\cdot\khat_1$ and $\cos\vartheta_4= -\khat_4\cdot\khat_2$. Then, one can check
\bes
\label{Pijexp}
\frac{1}{4}\det(P_i\cdot P_j) = {\bar \tau}^2/\eta^4 -  \khat_{4,\perp}^2 \sin^2\vartheta_3 \left[1+ {\cal O}(\tauhat^2/\eta^4)\right] + {\cal O}[({\hat\tau}/\eta^2)^4]
\ees
with
\bes
{\bar \tau} = \left(\frac{\tauhat_2}{2} - \tauhat_4\right) \sin\vartheta_3 + \frac{\tauhat_2}{2}\sin(\vartheta_3-\vartheta_4) + \left(\frac{\tauhat_2}{2}-\tauhat_3\right)\sin \vartheta_4\ . \label{taubar}
\ees

The other $P_i\cdot P_j$ terms in both $\rho$, eq.~(\ref{rhodef}), and $A_{CFT}$, eq.~(\ref{ACFTdef}), can likewise be expanded about $(\tau_{0i}, -\khat_i)$, but subleading terms enter the final expression at the same order as the neglected terms in (\ref{Pijexp}), and in particular their $\tauhat$ dependence contributes subleading corrections to the singularity.   Thus, (\ref{intermexp}) becomes 
\bes
\label{Intermexp}
\RHS = 2\pi\delta(\Rhat\sum_i\omega_i){\cal B}\lim_{\eta\rightarrow\infty}{{\cal L}\over \eta^8} (\Rhat \eta^2)^{5-d-2j} \int  {d\tauhat_2 d\tauhat_3d\tauhat_4 e^{-i\Rhat (\omega_2\tauhat_2+\omega_3\tauhat_3+\omega_4\tauhat_4)}\over( \khat_{4,\perp}^2 \sin^2\vartheta_3 -  {\bar \tau}^2/\eta^4 + 
\cdots)^\beta}
\ees
with
\begin{eqnarray}
{\cal B} &=&   g^2 e^{-i\pi (\Delta_1+\Delta_2)} \frac{C_{\Delta_1}C_{\Delta_2}}{ 2^{ \Delta_1+\Delta_2}}  {\cal F}(\sigma) \\
&&\left({-k_1\cdot k_3\over \omega_1\omega_3}\right)^{\beta-\Delta_1}  \left({-k_2\cdot k_4\over \omega_2\omega_4}\right)^{\beta-\Delta_2} \left({-k_1\cdot k_2\over \omega_1\omega_2}\right)^{\beta}  \left({-k_3\cdot k_4\over \omega_3\omega_4}\right)^{\beta} \ .\nonumber
 \end{eqnarray}

We can now change integration variables from $\tauhat_2$ to ${\bar \tau}$.  The leading singularity is then independent of $\tauhat_3$, $\tauhat_4$, and thus integrals over these give delta functions.  These, together with the energy-conserving delta function, enforce conservation of energy and momentum in the plane defined by $\khat_1$, $\khat_3$.  In particular, in the center-of-mass frame, $\omega_1=\omega_2$, we thus find $\vartheta_3=\vartheta_4 =\Theta$.

Collecting all powers of $\eta$ in (\ref{Intermexp}), together with 
the integral over $\bar \tau$, gives an expression of the form
\bes
\lim_{\eta\rightarrow\infty} \eta^{2(2\beta-2\Delta_1-2\Delta_2-2j+d+3)} \int d{\bar \tau} {e^{-i\Rhat \omega_2 {\bar\tau}/(\sin\Theta)}\over  (\eta^4 \khat_{4,\perp}^2 \sin^2\Theta - {\bar\tau}^2 )^\beta   }
\ees
As shown in appendix \ref{Nbeta}, this produces the appropriate delta function on transverse momenta precisely if 
\begin{eqnarray}
\beta= \Delta_1+\Delta_2+j -5/2\, ,\label{beta}
 \end{eqnarray}
and as long as $2\beta >d-2$.
 We will check in specific examples that this relation holds.
Then,
\bes
\lim_{\eta\rightarrow\infty} \eta^{2(d-2)} \int d{\bar \tau} {e^{-i\Rhat \omega_2 {\bar\tau}/(\sin\Theta)}\over  (\eta^4 \khat_{4,\perp}^2 \sin^2\Theta - {\bar\tau}^2)^\beta   }= {(\Rhat \omega_2)^{2\beta-d+1}  {\cal N}_\beta\over (\sin\Theta)^{2\beta-1}} \delta^{d-2}(\khat_{4\perp})
\ees
where the coefficient ${\cal N}_\beta$ is derived in appendix \ref{Nbeta}:
\bes
 \mathcal{N}_\beta=\frac{\pi ^{\frac{d+1}{2}} }{2^{2\beta-d}\Gamma \left( \beta \right) \Gamma\left( \frac{2\beta+3-d}{2} \right)}\ .
 \ees
 Combining the various factors, we then find
 \bes
 \RHS= (2\pi)^{d+1} \delta^{d+1}(\sum_i k_i) \mathcal{K}\,
 g^2 s^{j-1} \Big(\frac{-t}{s}\Big) ^{j-2} \Big(\frac{-u}{s}\Big) ^{3-j-\Delta_1-\Delta_2}
  \mathcal{F}\Big(\frac{-t}{s}\Big)  \ ,
 \ees
 where we have rewritten quantities in terms of the Mandelstam parameters (\ref{mandelstam}) and 
  \begin{eqnarray}
\mathcal{K}& =&
\frac{e^{- i \pi  (\Delta_1 + \Delta_2)} \mathcal{N}_\beta C_{\Delta_1} C_{\Delta_2}}{ 2(2\pi)^{d-2} \mathcal{D}_{\Delta_1}^2\mathcal{D}_{\Delta_2}^2 }\ . 
\end{eqnarray}
Thus, with appropriate singularity at $z={\bar z}$, the conformal field theory can reproduce the proper bulk kinematical structure.  Finally, comparing the two sides of (\ref{mainformulaa}), yields a general proposal for the form of the bulk reduced transition matrix element, in terms of the coefficient of the singularity:
 \begin{equation}
iT(s,t)=\mathcal{K}\,
 g^2 s^{j-1} \Big(\frac{-t}{s}\Big) ^{j-2} \Big(\frac{-u}{s}\Big) ^{3-j-\Delta_1-\Delta_2}
  \mathcal{F}\Big(\frac{-t}{s}\Big)
\ ,   \label{result}
\end{equation}
where $\cal F$ was defined in (\ref{limA}), and the constant $\cal K$ is given by 
\begin{eqnarray}
\mathcal{K}
=\frac{\pi^{\frac{d-3}{2}}\Gamma \left(\Delta_1  \right) \Gamma \left(\Delta_2  \right) 
\Gamma \left(\Delta_1 - \frac{d}{2} + 1 \right) \Gamma \left(\Delta_2 - \frac{d}{2} + 1 \right) }
{4^{j-2} \Gamma \left(\Delta_1 +\Delta_2+j- \frac{5}{2} \right) 
\Gamma \left(\Delta_1 +\Delta_2+j-1- \frac{d}{2} \right)}\ .
\end{eqnarray}

\subsection{Boundary kinematics of the singularity}

Clearly the singularity at $z={\bar z}$ is an essential feature of the boundary CFT, if it is going to reproduce the full bulk energy-momentum conservation.  In this subsection we investigate more closely the limit in which it is produced; then in the next section we will examine explicit correlators that exhibit this singularity.

The causal relations between the boundary points used in the previous section were: points 1 and 2 and points 3 and 4 were spacelike related
and points 3 and 4 lies inside the future lightcone of both points 1 and 2.  With these causal relations, the singularity is expected at $z={\bar z}$.  
From the bulk point of view, it is natural to consider the space of all points in AdS that are null related to the four boundary points  $b_i=(\tau_i,\bolde_i)$,
\begin{equation} 
X \in  \mathbb{R}^ {2,d}\ ,\ \ \ \ \ \ X\cdot P_i=0\ ,\ \ \ \ \ \ 
X^2=-R^2\ .
\end{equation} 
In general, this is an empty set. Indeed, the conditions $X\cdot P_i=0$ for generic $P_i$
 imply $X\in  \mathbb{R}^ {d-2}$ which is incompatible with $X^2=-R^2$.
Furthermore, the same statement applies to boundary points.

However, the condition of equal cross ratios $z=\bar{z}$ was seen  equivalent to
\begin{equation} 
{\rm Det} \, P_i\cdot P_j =0 \ , \label{det=0}
\end{equation} 
where the determinant is taken over the indices $i,j=1,\dots,4$.
This condition means that the four points $ P_i \in \mathbb{R}^ {2,d}$ are either linearly dependent
or they generate a null 4 dimensional submanifold.
In the latter case, it is convenient to write 
\begin{equation} 
\mathbb{R}^ {2,d}=\mathbb{R}^ {d-3} \times \mathbb{R}_ {\bar N} \times \left( \mathbb{R}_ {N} \times \mathbb{M}^ {3} \right)\ ,
\end{equation} 
where $\mathbb{R}_ {{\bar{N}}} \times  \mathbb{R}_ {N}=\mathbb{M}^2 $ is a split along two null directions $N, {\bar N}$, and
the factor in brackets represents the submanifold generated by the $P_i$'s.
For $X$ to be  lightlike related to all external points $P_i$, we need $X \in \mathbb{R}^ {d-3} \times \mathbb{R}_ {N}$, but
this is incompatible with  $X^2=-R^2$. We conclude
that there are no bulk points lightlike related to all external points.
Moreover, only the boundary point  $N \in \mathbb{R}_N$ is  lightlike related to all external points.

\begin{figure}
\centering
\includegraphics[height=6cm]{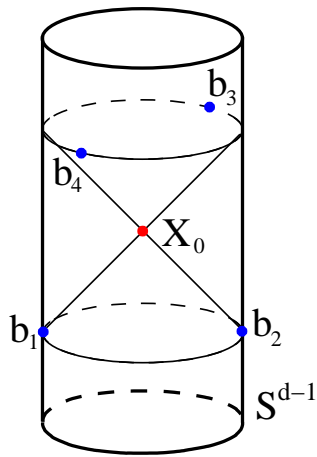}
\caption{Sketch of the boundary points configuration in AdS$_{d+1}$ for the Lorentzian kinematical condition of equal cross ratios $z=\bar{z}$.  Here, all such points are lightlike related to the bulk point $x_0$.
}
\label{CFTkin}
\end{figure}

In the degenerate case that we consider, where the external points are linearly dependent, 
the space orthogonal to all of them is $\mathbb{M}^ {d-1}$.
Then, the condition $X^2=-R^2$ defines a $(d-2)$-dimensional hyperboloid.
The space of all boundary points that are null related to the four points  $P_i$ 
consists of a $(d-3)$-sphere, which is the boundary of the $(d-2)$-dimensional hyperboloid in the bulk.

In the particular case of $d=2$ there is no null 4 dimensional submanifold of $\mathbb{R}^{2,2}$. Therefore, condition (\ref{det=0}) implies that the $P_i$'s are linearly dependent and we fall in the degenerate case described in the previous paragraph. Then, the $(d-2)$-dimensional hyperboloid in the bulk consists of a single point, which we can take to be our reference point $x_0$ as shown in figure \ref{CFTkin}). Unfortunately, we are unable to draw the more general higher dimensional cases where there are boundary points null related to all external points.

We conclude that the divergence of the four point function when $\bar{z} \to z$ would arise when there is
 a boundary point that is null related
to all the four external points of the correlation function.

\section{Examples}
\label{examples}
We shall now illustrate the appearance of the $z={\bar z}$ singularity and the application of our main result (\ref{result}) in some particular examples.
More precisely, we shall consider several explicit boundary four point functions (originally derived via euclidean bulk supergravity  tree computations), study their $\bar{z} \to z$ limit, and extract from this the corresponding bulk reduced transition matrix elements.
These will be found to  have precisely the correct form corresponding to the 
tree level interaction in flat space.

\subsection{Analytic continuation}
\label{analy}
We first describe the analytic continuation necessary to go from euclidean correlators to the lorentzian ones that we require.  In the euclidean regime, $z$ and ${\bar z}$ are indeed complex conjugate.  We find the lorentzian correlators by following the complex paths of $z$ and $\bar{z}$ generated by the appropriate Wick rotation.

The continuation path is described by the Wick rotation of AdS global time $\tau \to -i \tau e^{i \alpha}$ where $\alpha=0$ is the Euclidean regime and  $\alpha=\frac{\pi}{2}$ is the Lorentzian one.  The formula (\ref{boundcoords}) then gives
\bes
P\rightarrow(\cos(-i \tau e^{i \alpha}),\sin(-i \tau e^{i \alpha}),\bolde)\ .
\ees
With the kinematics described in section \ref{flatS}, this then yields the continuations
\begin{equation} 
\label{cplxpaths}
z=\cos^2 \frac{\Theta-i\pi e^{i \alpha}}{2} \ , \hspace{30pt}\bar{z}=\cos^2  \frac{\Theta+i\pi e^{i \alpha}}{2}\ .
\end{equation} 

\begin{figure} 
 \centering
\includegraphics[width=7cm]{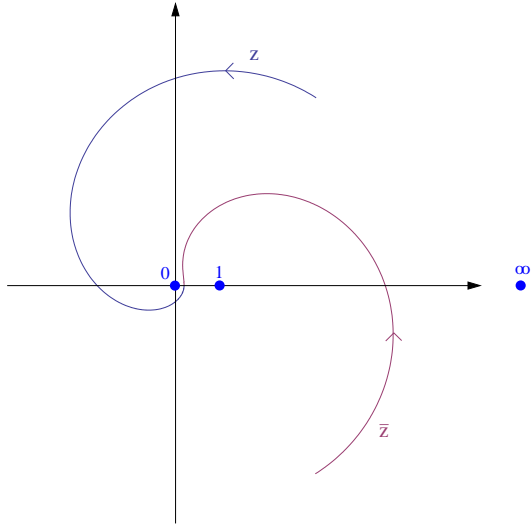}
\caption{ Complex paths $z(\alpha)$ and $\bar{z}(\alpha)$ starting from the Euclidean regime at  $\alpha=0$ to the Lorentzian  one at $\alpha=\frac{\pi}{2}$, for the particular scattering angle $\Theta=1$.
}\label{Wickrotation}
\end{figure}

In general the four point function is a multivalued function with branch points at $z,\bar{z}=0,1,\infty$. Therefore, it is important to evaluate the four point function in the appropriate Riemman sheet. The standard choice for the Euclidean four point function is to choose the branch cuts along the positive real axis. From figure \ref{Wickrotation} we see that under the Wick rotation $\bar{z}$ crosses this branch cut.  
It is also important that $z$ approaches the real axis from below and $\bar{z}$ from above
\begin{equation} 
z\to z-i\epsilon\ ,\hspace{25pt} \bar{z}\to \bar{z} +i\epsilon\ .
\end{equation}

\subsection{Contact interactions}

We start by considering a contact interaction with coupling $g^2$ between  our two scalar fields produced by a quartic vertex in AdS$_{d+1}$ . The coupling  $g^2$ has length dimension $d-3$ and therefore corresponds to $j=1$ in (\ref{limA}). Moreover, the tree level Witten diagram is simply given by 
\begin{equation} 
A_{CFT}(b_i)= g^2 R^{3-d}  \pi^{\frac{d}{2}}C_{\Delta_1}^2 C_{\Delta_2}^2 D_{\Delta_1 \Delta_2 \Delta_1 \Delta_2}(P_i)\ ,
\end{equation}  
where $D_{\Delta_i}$ is the standard D-function reviewed in appendix \ref{Dfunctions}.
Using equation (\ref{DtoDbar}) we find the reduced amplitude 
 \begin{equation} 
\mathcal{A}(z,\bar{z})= g^2 R^{3-d} \frac{\pi^{\frac{d}{2}} C_{\Delta_1} C_{\Delta_2}
 \Gamma\left(\Delta_1+\Delta_2-\frac{d}{2}\right)}
{2\Gamma^2(\Delta_1) \Gamma^2(\Delta_2)}
\bar{D}_{\Delta_1 \Delta_2 \Delta_1 \Delta_2}(u,v)\  ,
\end{equation}  
where $u$ and $v$ are defined in terms of $z$ and $\bar z$ in eq.~(\ref{uvdef}).
When $\Delta_1$ and $\Delta_2$ are positive integers, we can determine the small $\rho$ behavior of the $\bar{D}$-function using the techniques explained in appendix \ref{Dfunctions}. We obtain
\begin{equation} 
\bar{D}_{\Delta_1\, \Delta_2\, \Delta_1\,\Delta_2 }(u,v)\approx  2i  \pi^{\frac{3}{2}} 
 \Gamma\left(\Delta_1+\Delta_2-\frac{3}{2}\right) 
    \frac{\sigma (1-\sigma)^{\Delta_1+\Delta_2-2}}{(-\rho^2)^{\beta}} \ ,
\end{equation}  
with
\bes
\beta=\Delta_1+\Delta_2-\frac{3}{2}\ ,
\ees
in agreement with the prediction (\ref{beta}),
and which gives
\bes
\mathcal{F}(\sigma)=i\frac{ \pi^{\frac{d+3}{2}} C_{\Delta_1} C_{\Delta_2}
 \Gamma\left(\Delta_1+\Delta_2-\frac{d}{2}\right)\Gamma\left(\Delta_1+\Delta_2-\frac{3}{2}\right) }
{\Gamma^2(\Delta_1) \Gamma^2(\Delta_2)} 
 \sigma (1-\sigma)^{\Delta_1+\Delta_2-2}\ .
\ees
Inserting this expression into (\ref{result}) we obtain the simple reduced transition matrix element
\bes
T(s,t)=g^2\ ,
\ees
as expected for a contact interaction.

\subsection{Scalar exchange}

 We now consider scalar exchanges in AdS$_{d+1}$, which correspond to $j=0$.
As explained in \cite{D'Hoker:1999pj, D'Hoker:1999ni} , the associated four point function can be reduced to a finite sum of D-functions if
$2\Delta_1$ minus the conformal dimension of the t-channel exchanged scalar is a positive even integer.
We shall consider this particular case.
In appendix \ref{Dfunctions}, we show that different D-functions have different singular behavior at $\rho=0$.
In particular, the singularity at $\rho=0$ gets stronger as the sum of the indices of the D-function increases.  
Therefore, in the sum of D-functions obtained in  \cite{D'Hoker:1999ni} it is enough to keep
\begin{equation} 
A(b_i)\approx g^2 R^{5-d}\frac{\pi^{\frac{d}{2}} C_{\Delta_1}^2C_{\Delta_2}^2}{4(\Delta_1-1)^2 }\frac{1}{(-2P_1\cdot P_3)}
D_{\Delta_1-1\, \Delta_2 \, \Delta_1-1\, \Delta_2}(P_i)\ ,
\end{equation}  
which gives 
\begin{equation} 
\mathcal{A}(z,\bar{z})=g^2 R^{5-d}\frac{\pi^{\frac{d}{2}} C_{\Delta_1}C_{\Delta_2}\Gamma\left(\Delta_1+\Delta_2-1-\frac{d}{2}\right)}{8\Gamma^2(\Delta_1) \Gamma^2(\Delta_2)  }
\bar{D}_{\Delta_1-1\, \Delta_2 \, \Delta_1-1\, \Delta_2}(u,v)\ .
\end{equation}  
Using again the result
  \begin{equation} 
\bar{D}_{\Delta_1-1\, \Delta_2\, \Delta_1-1\,\Delta_2 }(u,v)\approx  2i \pi^{\frac{3}{2}} 
 \Gamma\left(\Delta_1+\Delta_2-\frac{5}{2}\right) 
    \frac{\sigma (1-\sigma)^{\Delta_1+\Delta_2-3}}{(-\rho^2)^{\Delta_1+\Delta_2-\frac{5}{2}}} \ ,
\end{equation}  
we confirm the predicted power of the singularity at $\rho=0$ and obtain
\begin{equation} 
\mathcal{F}(\sigma)=i \frac{\pi^{\frac{d+3}{2}} C_{\Delta_1}C_{\Delta_2}\Gamma\left(\Delta_1+\Delta_2-1-\frac{d}{2}\right) \Gamma\left(\Delta_1+\Delta_2-\frac{5}{2}\right) }{4\Gamma^2(\Delta_1) \Gamma^2(\Delta_2)  } \sigma (1-\sigma)^{\Delta_1+\Delta_2-3}\ .
\end{equation}  
The prescription (\ref{result}) then gives
\begin{equation}
T(s,t)= \frac{g^2}{ -t }\ ,
\end{equation}
which  agrees with the expected flat space result.

\subsection{Graviton exchange}

In \cite{D'Hoker:1999pj,D'Hoker:1999jp} the contribution to the four point function of $\Delta=4$ scalar operators from  t-channel graviton exchange in AdS$_5$ was determined.
In our conventions, the result reads\footnote{Since in our conventions $\mathcal{A}=1$ corresponds to the disconnected contribution, the normalization can be read of directly from equation (C.11) of \cite{D'Hoker:1999jp}.
 }
\begin{align}
\mathcal{A}=&\frac{2 G_5 }{3 \pi R^3} \left[  45 \bar{D}_{4 4 4 4} 
-4 \bar{D}_{1 4 1 4} - 20 \bar{D}_{2 4 2 4} - 23 \bar{D}_{3 4 3 4} \right. 
\\
&\hspace{1cm} \left.  + 15\frac{2-z-\bar{z}}{z\bar{z}}\bar{D}_{4 5 4 5}+\frac{2-z-\bar{z}+z\bar{z}}{z\bar{z}} \left(12 \bar{D}_{2 5 2 5}  + 
20 \bar{D}_{3 5 3 5} \right)  \right] \ ,\nonumber
 \end{align}
 where $G_5$ is the 5-dimensional Newton constant.
The leading singularity as $\rho \to 0$ comes from 
\begin{equation} 
\bar{D}_{4 5 4 5 }\left( u,v \right) \approx 2i
\pi^{\frac{3}{2}}  \Gamma\left(\frac{15}{2} \right)
\frac{\sigma (1-\sigma)^{7}}{(-\rho^2)^{\frac{15}{2}}}\ ,
\end{equation}  
computed in appendix \ref{Dfunctions}.
This gives
\begin{equation} 
\mathcal{A}\approx i  G_5 R^{-3} 40 \sqrt{\pi}\, \Gamma\left(\frac{15}{2}\right)
 \frac{(1-\sigma)^{8}}{\sigma}  \frac{1}{(-\rho^2)^{\frac{15}{2}}}\ ,
\end{equation}  
which has the predicted form (\ref{limA}). 
Using the value of the constant
\begin{equation} 
\mathcal{K}=\frac{ \sqrt{\pi} }{5\Gamma\left(\frac{15}{2} \right) }
\end{equation}  
and our main result (\ref{result}), one obtains the matrix element
\begin{equation} 
T(s,t) = 8 \pi G_5 s \,\frac{1-\sigma}{\sigma} =   8 \pi G_5 \frac{ s^2+t s }{-t}\ .
\end{equation}  
This agrees with the matrix element found in \cite{Barker:1966zz} for t-channel graviton exchange between minimally coupled massless scalars.

%


\section{Conclusion and open questions}

Since the AdS/CFT correspondence was first proposed, an important open question has been how to ``decode the hologram,'' that is, read off local bulk physics, particularly on scales short as compared to the AdS scale, from the boundary theory.  In this paper, we have suggested a partial answer to this question, for certain S-matrix elements.  In particular, we have argued that if the boundary CFT has a particular singularity structure, (\ref{limA}), with a characteristic leading behavior at $z={\bar z}$, then this suffices to produce important kinematical structure, in particular the {\it bulk} momentum conserving delta function.  Moreover, where the CFT does have such a singularity, the coefficient function of the singularity is expected to provide the reduced transition matrix element, as seen in (\ref{result}).

Moreover, we have seen this construction in operation, in the examples of section \ref{examples}.  There, we explicitly found that for certain ``CFT correlators,'' we could indeed reproduce the expected $T$-matrix elements. 

The reason quotes have been added to this last statement is that the correlators we have considered are, of course, correlators computed from the bulk supergravity, and not derived directly from an actual boundary conformal field theory.  This construction thus explains how such information {\it could} be encoded in and extracted from actual CFT correlators.  A very important question for the future is whether correlators computed from actual boundary CFTs have the appropriate structure.  Thus, the present construction provides an important test for CFTs, which can be used to determine whether they encode properties of a bulk local theory.  In some respects this seems a non-trivial test, as it requires a very precise fine-grained structure exhibited in the $\eta\rightarrow\infty$ limit, so in correlators, in the $z\rightarrow {\bar z}$ limit, probing very short scales.  It will be interesting to see in what cases such structure is produced in bona-fide conformal field theories.

While we view this as an important test for CFTs, it is not a complete one. For example, the T-matrix of a bulk theory that is at least approximately local on scales long as compared to the string or Planck scale is expected to have certain other properties, such as characteristic growth at high energies.   Moreover, a complete reconstruction of the S-matrix would require that one can recover other S-matrix elements, for example outside the plane-wave limit\cite{MGSG}, and for multi-particle processes.  Other related investigations include studying processes with external particles with spin, and examining the structure of loop and string amplitudes.  
The inclusion of string and loop effects introduces additional parameters in the correlator, namely $\ell_s/R$ and $\ell_{Pl}/R$. 
We then expect that the nature of the  $z=\bar{z}$ singularity of the correlator to change as 
 $z-\bar{z}$ becomes smaller relative to these parameters.
However, since the singularity has encoded the overall momentum-conserving delta function, one expects aspects of the structure we found in this paper to remain valid for amplitudes at higher-orders, or even non-perturbatively.

In short, these methods suggest a way that candidate CFTs could be probed for anticipated bulk structures.  Given a candidate CFT, one might  investigate the behavior of its correlators for $z\approx \bar{z}$ to see whether they have the correct structure to encode various  bulk phenomena, such as loop effects, string excitations, and small black holes.
\vskip .15in

\noindent{\bf Acknowledgements} We wish to thank L. Cornalba, M. Costa, T. Okuda, E. Witten, and especially J. Polchinski for discussions.  
 MG and SBG gratefully acknowledge the kind hospitality of the CERN theory group, where part of this work was carried out.   
 The work of MG and SBG was supported in
part by the U.S. Dept. of Energy under Contract
DE-FG02-91ER40618, and by grant RFPI-06-18 from the Foundational
Questions Institute (fqxi.org).  
MG is supported by a Marie Curie Early Stage Research
Training Fellowship of the European Community's Sixth Framework Programme
under contract number MEST-2005-020238-EUROTHEPHY.
JP is funded by the FCT fellowship SFRH/BPD/34052/2006, partially  by the grant
CERN/FP/83508/2008, and supported in part by the National Science Foundation under Grant No. NSFPHY05-51164.

\appendices

\section{The transverse delta function} \label{Nbeta}

In this section, we verify the formula used in section \ref{CFTconstruction} for the transverse delta function,
\bes
 {\cal N}_\beta \delta^{n}({\vec \kappa})= \lim_{\eta\rightarrow\infty} \int d\nu e^{-i\nu} {\eta^{2n}\over \left[\eta^4\kappa^2 -(\nu+i\epsilon)^2\right]^\beta} \ .
 \ees
The $i\epsilon$ prescription was obtained from the Wick rotation of AdS global time explained in section \ref{analy}.
In particular, we take $\tau \to \tau(1-i\epsilon)$ which gives $\tauhat_2 \to \tauhat_2 +i\epsilon$, $\tauhat_3 \to \tauhat_3 -i\epsilon$ and $\tauhat_4 \to \tauhat_4-i\epsilon$. Equation (\ref{taubar}) then gives the final prescription $\bar{\tau} \to \bar{\tau} +i\epsilon$.

First, we note that for $\kappa^2\neq0$, the function vanishes in the limit, as long as $2\beta>n$.
Next, let us compute the integral of  this expression over $n$-dimensional $\kappa$ space.  We begin with 
\bes
{\cal N}_\beta =  \lim_{\eta\rightarrow\infty}\int d^n\kappa \int d\nu e^{-i\nu} {\eta^{2n} \over  \left[\eta^4\kappa^2 -(\nu+i\epsilon)^2\right]^\beta}\ 
\ees
The quantity $\eta$ scales out trivially.  Then, we can rewrite
\bes
{\cal N}_\beta = \int d^n\kappa \int_0^\infty d\nu \left[ {e^{-i\nu}\over (\kappa^2 -\nu^2 -i\epsilon)^\beta} + c.c.\right]
\ees
where {\it c.c.} denotes the hermitian conjugate. 
 The denominator can be exponentiated by the Schwinger trick, to yield
\bes
{\cal N}_\beta = {1\over \Gamma(\beta)} \int d^n\kappa \int_0^\infty d\nu e^{-i\nu} \int_0^\infty id\zeta (i\zeta)^{\beta-1} e^{-i\zeta(\kappa^2 -\nu^2 -i\epsilon)} + c.c.\ 
\ees
Then, one does the gaussian integral over $\kappa$ to find 
\bes
{\cal N}_\beta = {\pi^{n/2}\over \Gamma(\beta)}  \int_0^\infty id\zeta (i\zeta)^{\beta-n/2-1} \int_0^\infty d\nu e^{i\zeta\nu^2 -i\nu -\epsilon\zeta} + c.c.\ 
\ees
We can now rotate $\nu \to e^{-i\frac{\pi}{2}}\nu$ and $\zeta \to e^{i\frac{3\pi}{2}}\zeta$,
\begin{eqnarray}
 \mathcal{N}_\beta 
&=& -i e^{i\pi (2\beta-n)} \frac{\pi^{\frac{n}{2}}}{\Gamma(\beta)}
\int_0^\infty d\nu  \int_0^\infty d\zeta \zeta^{\beta-1-\frac{n}{2}}
e^{-\zeta  \nu^2  - \nu  } +c.c.\\
&=& -i e^{i\pi (2\beta-n)} \frac{\pi^{\frac{n}{2}} \Gamma\left(\beta-\frac{n}{2}\right)}{\Gamma(\beta)}
\int_0^\infty d\nu   \nu^{n-2\beta}
e^{ - \nu  } +c.c.\\
&=& -i e^{i\pi (2\beta-n)} \frac{\pi^{\frac{n}{2}} \Gamma\left(\beta-\frac{n}{2}\right)\Gamma\left(n+1-2\beta\right)}{\Gamma(\beta)} +c.c.\\
&=& 2\sin\pi (2\beta-n) \frac{\pi^{\frac{n}{2}} \Gamma\left(\beta-\frac{n}{2}\right)\Gamma\left(n+1-2\beta\right)}{\Gamma(\beta)} \\
&=&  \frac{2\pi^{\frac{n+2}{2}} \Gamma\left(\beta-\frac{n}{2}\right)}{\Gamma(\beta)\Gamma\left(2\beta-n\right)} \\
 &=&
 \frac{\pi^{\frac{n+3}{2}}  }{2^{2\beta-n-2}\Gamma(\beta)\Gamma\left(\beta -\frac{n-1}{2}\right)}\ .
\end{eqnarray}

\section{D--functions} \label{Dfunctions}

\subsection{Basics}

The D--functions are defined as integrals over hyperbolic space \cite{D'Hoker:1999pj,Dolan:2000ut}, 
\be
D_{\Delta_{i}}^{d}\left(  P_{i}\right)  =\pi^{-\frac{d}{2}} \int_{H_{d+1}} dX
\,{\textstyle\prod\nolimits_{i}}
\,\left(  -2X\cdot P_{i}\right)^{-\Delta_{i}}\ ,
\ee
where the points $P_{i}$  are future directed  null vectors of the embedding space $\mathbb{M}^{d+2}$ of hyperbolic space $H_{d+1}$ and we set $R=1$.
Introducing Schwinger parameters one can derive the following integral representation 
\begin{align}
D_{\Delta_{i}}^{d}\left(  P_{i}\right)   & =
\frac{\Gamma\left( \Delta- \frac{d}{2}\right)  }{
{\textstyle\prod\nolimits_{i}}
\Gamma\left(  \Delta_{i}\right)  }\int_0^\infty
{\textstyle\prod\nolimits_{i}}\,
dt_{i}\,t_{i}^{\Delta_{i}-1}~e^{-\frac{1}{2}\sum_{i,j}t_{i}
t_{j}~P_{ij}}
\label{Dfunc}
\end{align}
where $P_{ij}=-2P_{i}\cdot P_{j}\ge 0$ and $\Delta=\frac{1}{2}{\textstyle\sum\nolimits_{i}}\Delta_{i}$.

The D--functions are invariant under Lorentz transformations of $\mathbb{M}^{d+2}$ and are homogeneous functions of $P_i$ with weight $-\Delta_i$. Therefore they can be reduced to functions of the invariant cross ratios,
\begin{equation}
 \frac{P_{ij}P_{kl} }{P_{ik} P_{jl}}\ .
\end{equation}
with $i\neq j\neq k\neq l$.
In particular, the four-point function can be written as
\begin{equation} 
D_{\Delta_{i}}^{d}\left(  P_{i}\right)   =
\frac{\Gamma\left(  \Delta-\frac{d}{2}\right)  }{2
{\textstyle\prod\nolimits_{i}}
\Gamma\left(  \Delta_{i}\right)  }
\frac{ \left(\frac{ P_{14}}{ P_{13}P_{34}}\right)^{\frac{\Delta_3-\Delta_1}{2}}
 \left(\frac{ P_{13}}{ P_{14}P_{34}}\right)^{\frac{\Delta_4-\Delta_2}{2}}}{ P_{13}^{\Delta_1} P_{24}^{\Delta_2} }
\bar{D}_{\Delta_{i}}\left( u,v \right)  \label{DtoDbar}
\end{equation}  
where $\bar{D}_{\Delta_{i}}$ is a function of the conformally invariant cross ratios
\begin{equation} 
\label{uvdef}
u=\frac{ P_{12}P_{34}}{P_{13} P_{24}}=\frac{1}{z\bar{z}}\ ,\hspace{15pt}
v=\frac{P_{14} P_{23}}{P_{13} P_{24}}=\frac{(1-z)(1-\bar{z})}{z\bar{z}}\ .
\end{equation}  
This function satisfies the following relations \cite{Dolan:2000ut}
\begin{align}
\bar{D}_{\Delta_1\, \Delta_2\, \Delta_3\,\Delta_4 }\left( u,v \right) 
&= -\partial_u \bar{D}_{\Delta_1-1\, \Delta_2-1\,\Delta_3\,\Delta_4}\left( u,v \right)  \nonumber \\
&= -\partial_v \bar{D}_{\Delta_1\, \Delta_2-1\,\Delta_3-1\,\Delta_4}\left( u,v \right)  \nonumber \\
&=  \bar{D}_{\Delta_3\, \Delta_2\,\Delta_1\,\Delta_4}\left( v,u \right)   \label{recrel} \\
&=  u^{\Delta_3+ \Delta_4- \Delta } 
 \bar{D}_{\Delta_4\, \Delta_3\,\Delta_2\,\Delta_1}\left( u,v \right)  \nonumber  \\
 &=  v^{\Delta_4- \Delta } 
 \bar{D}_{\Delta_2\, \Delta_1\,\Delta_3\,\Delta_4}\left( u/v,1/v \right) \nonumber
\end{align}
Finaly, we recall \cite{Dolan:2000uw,Dolan:2000ut} 
that the  function $\bar{D}_{1111}$ can be written explicitly as
\begin{eqnarray}
\bar{D}_{1111}  
 &=&\frac{ z\bar{z}}{ z -\bar{z}}\left[2 {\rm Li_2}(z)   -2 {\rm Li_2}(\bar{z})
+\log(z\bar{z}) \log\frac{1-z}{1-\bar{z}}\right]\ . \label{expD1111}
\end{eqnarray}

\subsection{Singular limit}

We shall now consider the singular limit, $z \to \bar{z}$, of some D-functions which we used in the main text.
We start by studying the function $\bar{D}_{1111}$. From the explicit expression (\ref{expD1111}) it is clear that  $\bar{D}_{1111}$ is regular when  $z \to \bar{z}$. This happens because the expression in square brackets in (\ref{expD1111}) vanishes as  $z \to \bar{z}$ and cancels the explicit pole in front. However, after the analytic continuation of figure \ref{Wickrotation} the function  $\bar{D}_{1111}$ has a real singularity.
To see this, let us place all the branch cuts of the expression in square brackets in (\ref{expD1111}) along the positive real axis,
\begin{equation}
2 {\rm Li_2}(z)   -2 {\rm Li_2}(\bar{z})
+\left(\log(-z)+\log(-\bar{z})\right)\left( \log(1-z)-\log(1-\bar{z})\right)\ .\label{expsqbra}
\end{equation}
Under the analytic continuation of figure \ref{Wickrotation}, $z$ does not cross any branch cut and $\bar{z}$ crosses all branch cuts, yielding the discontinuities
\begin{align}
 \log(-\bar{z}) &\to  \log(-\bar{z}) +2\pi i\\
 \log(1-\bar{z}) &\to  \log(1-\bar{z}) +2\pi i\\
 {\rm Li_2}(\bar{z}) &\to  {\rm Li_2}(\bar{z}) -2\pi i \log(\bar{z})
\end{align}
This turns (\ref{expsqbra}) into
\begin{eqnarray}
&&2 {\rm Li_2}(z)   -2 {\rm Li_2}(\bar{z}) + 4\pi i \log(\bar{z})  \label{expsqbralor}    \\
&+&\left(\log(-z)+\log(-\bar{z})+2\pi i \right)\left( \log(1-z)-\log(1-\bar{z}) -2\pi i\right) \nonumber
\end{eqnarray}
Following figure \ref{Wickrotation} we now take the  $\rho \to 0$ limit in the form
\begin{equation}
z\to \sigma e^{-\rho} -i\epsilon \ , \ \ \ \ \ \ \ \ \ \ 
\bar{z}\to \sigma e^{\rho} +i\epsilon \ .
\end{equation}
This drastically simplifies (\ref{expsqbralor}) to $4\pi^2$ and gives the small $\rho$ behavior
\begin{equation} 
 \bar{D}_{1111}  \approx   -\frac{2\pi^2 \sigma}{\rho}\ . 
\end{equation}  

It is now very easy to determine the small $\rho$ behavior of other D--functions with
positive and integer $\Delta$ and $\Delta_i$.
We just need to use the recursion relations (\ref{recrel}) and 
\begin{align}
\partial_u&=\frac{ z\bar{z}}{ z -\bar{z}}\left[ z(1-z)\partial_z -\bar{z}(1-\bar{z})\partial_{\bar{z}}\right]\\
&=-\frac{\sigma^3}{2}\partial_\sigma +\frac{\sigma(1-\sigma \cosh \rho)}{2 \sinh\rho} \partial_\rho 
\approx  \frac{\sigma(1-\sigma)}{2 \rho} \partial_\rho\ ,\\
\partial_v&=\frac{ z\bar{z}}{ z -\bar{z}}\left[\bar{z} \partial_{\bar{z}} - z\partial_z \right]
=-\frac{\sigma}{2 \sinh\rho} \partial_\rho 
\approx  - \frac{\sigma}{2 \rho} \partial_\rho\ .
\end{align}
For example,
\begin{align}
\bar{D}_{4 5 4 5 }\left( u,v \right) &= (-\partial_u)^3  \bar{D}_{1 2 4 5 }\left( u,v \right)\\
&= (-\partial_u)^3  u^3 \bar{D}_{5 4 2 1 }\left( u,v \right)\\
&= (-\partial_u)^3  u^3  (-\partial_u)^3 \bar{D}_{2 1 2 1 }\left( u,v \right)\\
&= (-\partial_u)^3  u^3  (-\partial_u)^3 v^{-2} \bar{D}_{1 2  2 1 }\left( u/v,1/v \right)\\
&= (-\partial_u)^3  u^3  (-\partial_u)^3 v^{-2} \left[ -\partial_v \bar{D}_{1 1  1 1 }\left( u,v \right) \right]_{u\to \frac{ u}{v}, v\to \frac{1}{v}}\ ,
\end{align}
which in the singular limit reduces to
\begin{align}
 \bar{D}_{4 5 4 5 }&\approx (1-\sigma)^2\sigma^4
 \left(-\frac{1}{2 \rho} \partial_\rho\right)^6  
\left[ -\frac{\sigma}{2 \rho} \partial_\rho  \frac{2\pi^2 \sigma}{\rho} \right]_{\sigma \to1-\sigma, \rho^2 \to \frac{\sigma^2 \rho^2}{(\sigma-1)^2}}  \\
 & \approx \pi^2 (1-\sigma)^7\sigma
 \left(-\frac{1}{2 \rho} \partial_\rho\right)^6  
 \frac{ 1}{(\rho^2)^{\frac{3}{2}}} \\
  & \approx 2 \pi^{\frac{3}{2}} \Gamma\left(\frac{15}{2}\right) 
 \frac{ (1-\sigma)^7\sigma}{(\rho^2)^{\frac{15}{2}}} \ . 
 \end{align}

The recursion relations (\ref{recrel}) either preserve $\Delta$ or increase it by one unit when taking derivatives $\partial_u$ or $\partial_v$. On the other hand, both $\partial_u$ and $\partial_v$ contribute a factor of $\rho^{-2}$ to the singularity at $\rho=0$. Therefore, a D--function with positive and integer $\Delta$ and $\Delta_i$ has the following small $\rho$ behavior,
\begin{equation} 
\bar{D}_{\Delta_1\, \Delta_2\, \Delta_3\,\Delta_4 } \sim \rho^{3-2\Delta}\ .
\end{equation}

Another particular example is
\begin{align}
 \bar{D}_{\Delta_1\, \Delta_1\, \Delta_2\,\Delta_2 } &=  
 (-\partial_u)^{\Delta_1-1}  u^{\Delta_2-1}  (-\partial_u)^{\Delta_2-1}   \bar{D}_{1 1  1 1 }\left( u,v \right)\\
   &\approx -2 \pi^{\frac{3}{2}}  \Gamma\left(\Delta_1+\Delta_2-\frac{3}{2}\right) 
    \frac{\sigma^{\Delta_1-\Delta_2+1} (1-\sigma)^{\Delta_1+\Delta_2-2}}{(\rho^2)^{\Delta_1+\Delta_2-\frac{3}{2}}} 
 \end{align}
for integer $\Delta_1$ and $\Delta_2$.
Using the relation
\begin{equation} 
\bar{D}_{\Delta_1\, \Delta_2\, \Delta_1\,\Delta_2 }(u,v)= v^{-\Delta_1} \bar{D}_{\Delta_1\, \Delta_1\, \Delta_2\,\Delta_2 }(1/v,u/v)\ ,
\end{equation}  
we find
\begin{equation} 
\bar{D}_{\Delta_1\, \Delta_2\, \Delta_1\,\Delta_2 }(u,v)\approx  -2 \pi^{\frac{3}{2}} 
 \Gamma\left(\Delta_1+\Delta_2-\frac{3}{2}\right) 
    \frac{\sigma (1-\sigma)^{\Delta_1+\Delta_2-2}}{(\rho^2)^{\Delta_1+\Delta_2-\frac{3}{2}}} \ .
\end{equation}

\end{document}